\documentclass[twocolumn,aps,prl,groupedaddress,showpacs]{revtex4-1}
\usepackage{amsfonts}
\usepackage{amsmath}
\usepackage{amssymb}
\usepackage{txfonts}
\usepackage{pxfonts}
\usepackage{graphicx,bm,units,yfonts}
\usepackage[table]{xcolor}

\newcommand{\ket}[1]{| #1 \rangle}
\newcommand{\bra}[1]{\langle #1 |}

\begin{document}

\title{Topological Criticality in the Chiral-Symmetric AIII Class at Strong Disorder}
\author{Ian Mondragon-Shem and Taylor L. Hughes}
\affiliation{Department of Physics, University of Illinois-Urbana Champaign, Urbana, IL 61801, USA}
\author{Juntao Song and Emil Prodan}
\affiliation{Department of Physics, Yeshiva University, New York, NY 10016, USA}

\begin{abstract}
The chiral AIII symmetry class in the periodic table of topological insulators contains topological phases classified by a winding number $\nu$ for each odd space-dimension. An open problem for this class is the characterization of the phases and phase-boundaries in the presence of strong disorder. In this work, we derive a covariant real-space formula for $\nu$ and, using an explicit 1-dimensional disordered topological model, we show that $\nu$ remains quantized and non-fluctuating when disorder is turned on, even though the bulk energy-spectrum is {\it completely localized}. Furthermore, $\nu$ remains robust even after the insulating gap is filled with localized states, but when the disorder is increased even further, an abrupt change of $\nu$ to a trivial value is observed. Using exact analytic calculations, we show that this marks a critical point where the localization length diverges. As such, in the presence of disorder, the AIII class displays markedly different physics from everything known to date, with robust invariants being carried entirely by localized states and bulk extended states emerging from an absolutely localized spectrum. Detailed maps and a clear physical description of the phases and phase boundaries are presented based on numerical and exact analytic calculations.
\end{abstract}


\maketitle

In the periodic classification table of topological insulators and superconductors \cite{schnyder2008, Ryu2010,Kitaev2009},  the unitary class A, which includes integer quantum Hall insulators, is arguably the most well-understood class, especially in the presence of disorder. Here, in even space-dimensions, the topological phases are characterized by the topologically invariant Chern numbers \cite{Qi:2008cg}. These topological integers are explicitly known to be robust against disorder \cite{Prodan2010ew, ProdanJPA2013vy}, that is, they cannot jump between different quantized values unless the Fermi level crosses a region of extended quantum states \cite{HalperinPRB1982er,bellissard1994}. Consequently, phases with different Chern numbers are necessarily separated by phase-boundaries carrying extended bulk states. Furthermore, the Chern numbers are known to be carried by bulk extended states that are embedded within a large set of localized states. When the disorder is increased, the systems in class A undergoe the ``levitation and annihilation" process where the bulk extended states above and below the Fermi level levitate toward one another in the energy spectrum and then annihilate upon collision, leading to topological phase transitions. This complete picture that we have for class A is very often assumed to apply to all symmetry classes in the periodic table since it has been observed in several different symmetry classes \cite{Onoda2007,Xu2012,gilbert2012,Leung2012}.

Looking over the periodic table, one immediately notices that the AIII chiral-unitary class is the natural complement of the A unitary class, but in odd space-dimensions \cite{schnyder2008,hosur2010}.
In this Letter, we demonstrate that the disorder-driven topological phase transitions in the AIII symmetry class can be strikingly different. For a generic 1-dimensional (1D) two-band model in AIII-class, we use well established methods to show that the bulk energy spectrum harbors no extended states. Yet, by using a covariant real-space representation of the integer-valued winding number $\nu$, which characterizes the different phases in the AIII-class, we find that $\nu$ remains quantized and non-fluctuating, even after the disorder completely fills the spectral gap with localized states. After increasing the disorder strength even further, a sharp transition is observed where $\nu$ drops from the topological to the trivial value. Using exact analytics, which are also confirmed by numerics, we demonstrate that the localization length of the disordered model diverges at this transition point. Our findings demonstrate that robust topological numbers (even integer-valued invariants) can be carried entirely by {\it localized states}, and that disorder can drive the {\it completely localized} topological phase through a {\it delocalized} critical point, in striking contrast to what has been observed so far for disordered topological insulators.

\begin{figure}
\includegraphics[width=8cm]{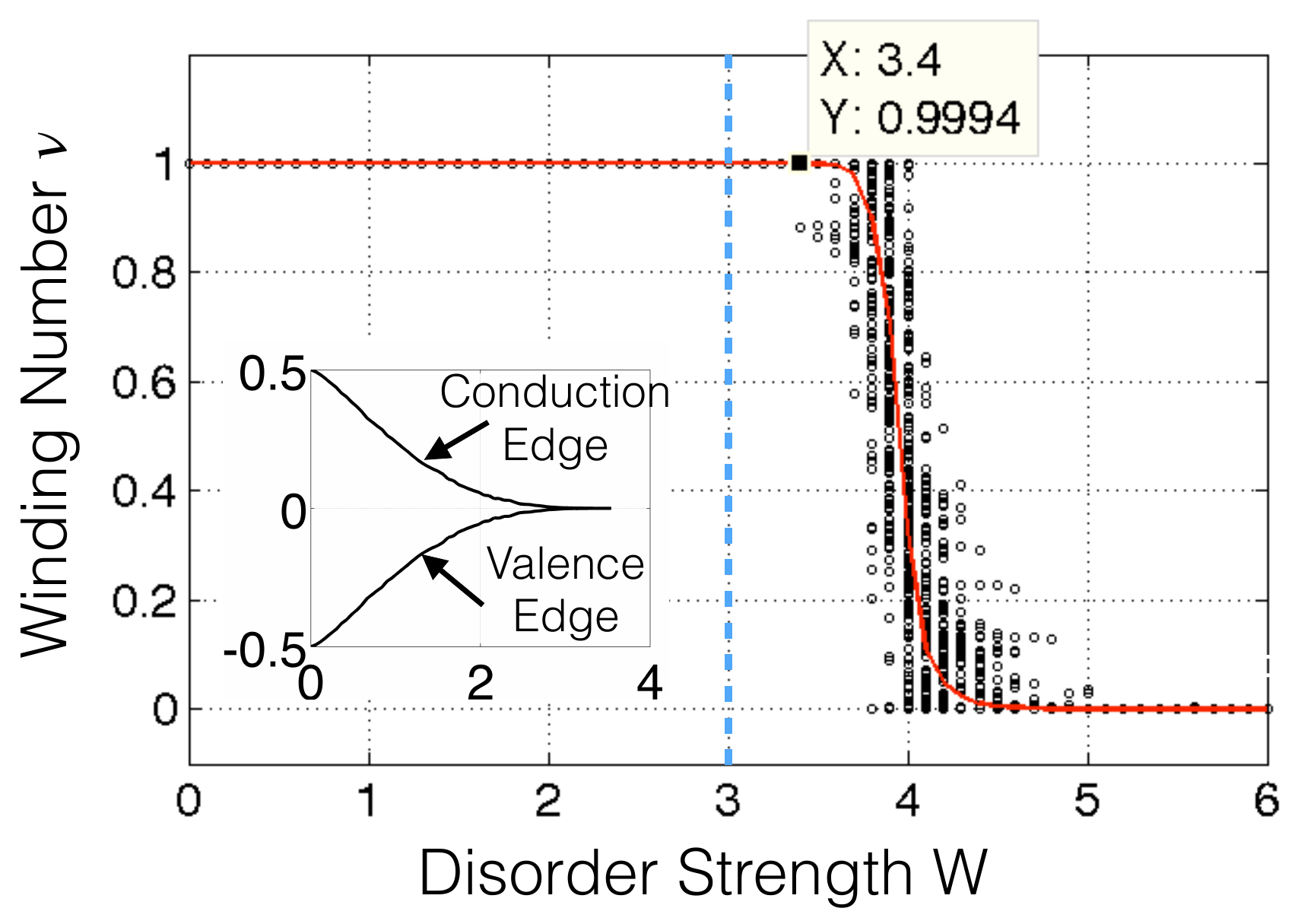}
\caption{(Color online) Evolution of the winding number $\nu$ (Eq.~\ref{RWindingNr}) with disorder: $W_1=0.5W$ and $W_2=W$. The raw, un-averaged data for 200 disorder configurations is shown by the scattered points and the average by the solid line. Inset: the conduction and valence edges as functions of W, indicating a spectral gap closing at $W \approx 3$ (marked by the dashed line in the main panel). The marked data point reports a quantized $\nu=0.9994$, at disorder well beyond $W=3$.}\label{InvariantVsW}
\end{figure}

Without further ado, we now present the analysis and the results. The disordered model we work with is:
\begin{equation}\label{Model}
H=\sum_{n} \left \{  t_n \left[\nicefrac{1}{2} \ c_n^\dagger ( \sigma_1+i \sigma_2 ) c_{n+1}+ \text{h.c.}\right]+ m_n \  c^{\dagger}_n \sigma_2  c_n  \right \},
\end{equation}
where $n$ runs over the lattice sites, $\sigma_\alpha$'s are the Pauli matrices and $c^\dagger_n=({c^\dagger _{n,A}},{c^\dagger_{n,B}})$ creates particles of orbital-type $A$ or $B$ at site $n$. The disorder is present on both the hopping and onsite potentials:
$$ t_n = 1 +W_1 \omega_n, \ \ m_n = m+W_2 \omega'_n,$$
where $\omega_n,\, \omega'_n$ are independent randomly generated numbers drawn from the uniform distribution $\left[-0.5,\,0.5\right]$.
The model in Eq.~\ref{Model}  preserves only the chiral symmetry $S H S^{-1}=-H$, with $S=\sum_n c^\dagger_n\sigma_3 c_n,$ as the last term in Eq.~\ref{Model} breaks both the particle-hole ($C=\sigma_3 K$) and time-reversal ($T=K$, $K$= complex conjugation) symmetries. Despite its simplicity, the behavior of this model will be representative for the 1D AIII class, because any gapped chiral-symmetric system can be adiabatically deformed into an independent sum of $2$-band models like Eq. \ref{Model}.

We begin with the analysis of the model in the clean limit $W_{1,2}=0$. The Bloch Hamiltonian takes the form:
\begin{equation}
h(k) = \left (
\begin{array}{cc}
0 & e^{ik} - i m \\
e^{-ik} +i m & 0
\end{array}
\right ),
\end{equation}
which obeys $\sigma_3h(k)\sigma_3^{-1}=-h(k)$. Being in class AIII, the topological invariant of the model is given by the winding number of the off-diagonal part of $h(k)$ \cite{Schnyder:2009qa}:
\begin{equation}\label{KWindingNr}
\nu = \frac{1}{2 \pi i}\int_0^{2\pi } dk \ (e^{ik} - i m)^{-1}\partial_k(e^{ik} - i m),
\end{equation}
from which  the residue theorem gives $\nu=+1$ for $m\in (-1,1)$ and $\nu=0$ otherwise. We note that the bulk energy-gap closes precisely at $m=\pm 1$, which signals the topological phase transitions in the clean system. In general, $\nu$ can take on any integer value and is gauge-invariant under a change of phase of the Bloch wavefunctions. If $n_{\pm}$ denote the numbers of bound states of each chirality at one end of an open chain, then topology enforces the bulk-edge correspondence: $\nu=\pm(n_{+}-n_{-}).$

One can find an additional physical interpretation of $\nu$ by noticing that it can  be generically re-written as a ``skew-polarization:"
\begin{equation}
\nu = \frac{1}{\pi}\int_0^{2\pi} dk \ \tilde{A}(k), \ \ \tilde{A}(k)= i\sum_{\alpha\in occ.} \langle S u_\alpha(k)|\partial_k |u_\alpha(k) \rangle,
\end{equation}\noindent where $|u_{\alpha}(k)\rangle$ is the Bloch function for band $\alpha.$ The skew-polarization is gauge-invariant precisely because $|u_{\alpha}(k)\rangle$ and $S|u_{\alpha}(k)\rangle$ are orthogonal. For 1D topological insulators, one is probably more familiar with the standard electric polarization \cite{king1993,ortiz1994,zak1989,QiPRB2008ng,hughes2011,turner2012}:
\begin{equation}
P = \frac{1}{2\pi}\int_0^{2\pi} dk \ A(k), \ \ A(k)= i\sum_{\alpha\in occ.} \langle  u_\alpha(k)|\partial_k |u_\alpha(k) \rangle,
\end{equation}
which is not gauge invariant but can change by an integer under gauge transformations. Since $ \langle  u_\alpha(k)|\partial_k |u_\alpha(k) \rangle= \langle  S u_\alpha(k)|\partial_k |S u_\alpha(k) \rangle$, the polarizations of the negative and positive energy bands are equal. And since the polarization of a system with all bands filled must be an integer, this implies that $2P\in \mathbb{Z}$, and thus $P$ is quantized in units of $1/2$ in class AIII. In the supplementary information, we show that $2P=\nu\mod 2$, which gives a new physical interpretation for $\nu$.

In 1D, there is no obstruction to defining localized Wannier functions from the occupied states in an insulating phase \cite{kivelson1982,marzari1997,thonhauser2006,Brouder:2007p233,hastings2010}, even if topological. In our case, setting $m=0$ places the system deep in the topological phase and the Bloch wavefunctions become $\vert u_{\pm}(k)\rangle=\tfrac{1}{\sqrt{2}}(1,\; \mp e^{ik})^{T}.$ From the set of occupied Bloch functions $\vert u_{-}(k)\rangle$, we can explicitly construct a set of ultra-localized Wannier functions near each site $n$: $W^{(-)}_{n}=\tfrac{1}{\sqrt{2}}(\vert n, B\rangle +\vert n+1, A\rangle)$, having weight only on two neighboring sites. When the system is tuned away from $m=0$, but still in the topological phase, the Wannier functions $W^{(-)}_{n}$ will gradually spread but still remain exponentially localized at the mid-bond between sites $n$ and $n+1$. It is then somewhat surprising that this system, and, in fact, all 1D topological insulator/superconductor phases can support a non-trivial integer topological invariant, even though the occupied space can be represented entirely using localized bulk states. Additionally, the well-known levitation-annihilation process for disorder-driven topological phase transitions in free-fermion topological phases cannot possibly apply here because there are no delocalized bulk modes which carry the topological invariant. Instead, each single-particle electron state carries part of the topological invariant, and because of this, a different type of disorder-driven transition must occur as we now discuss.

\begin{figure}
\includegraphics[width=8cm]{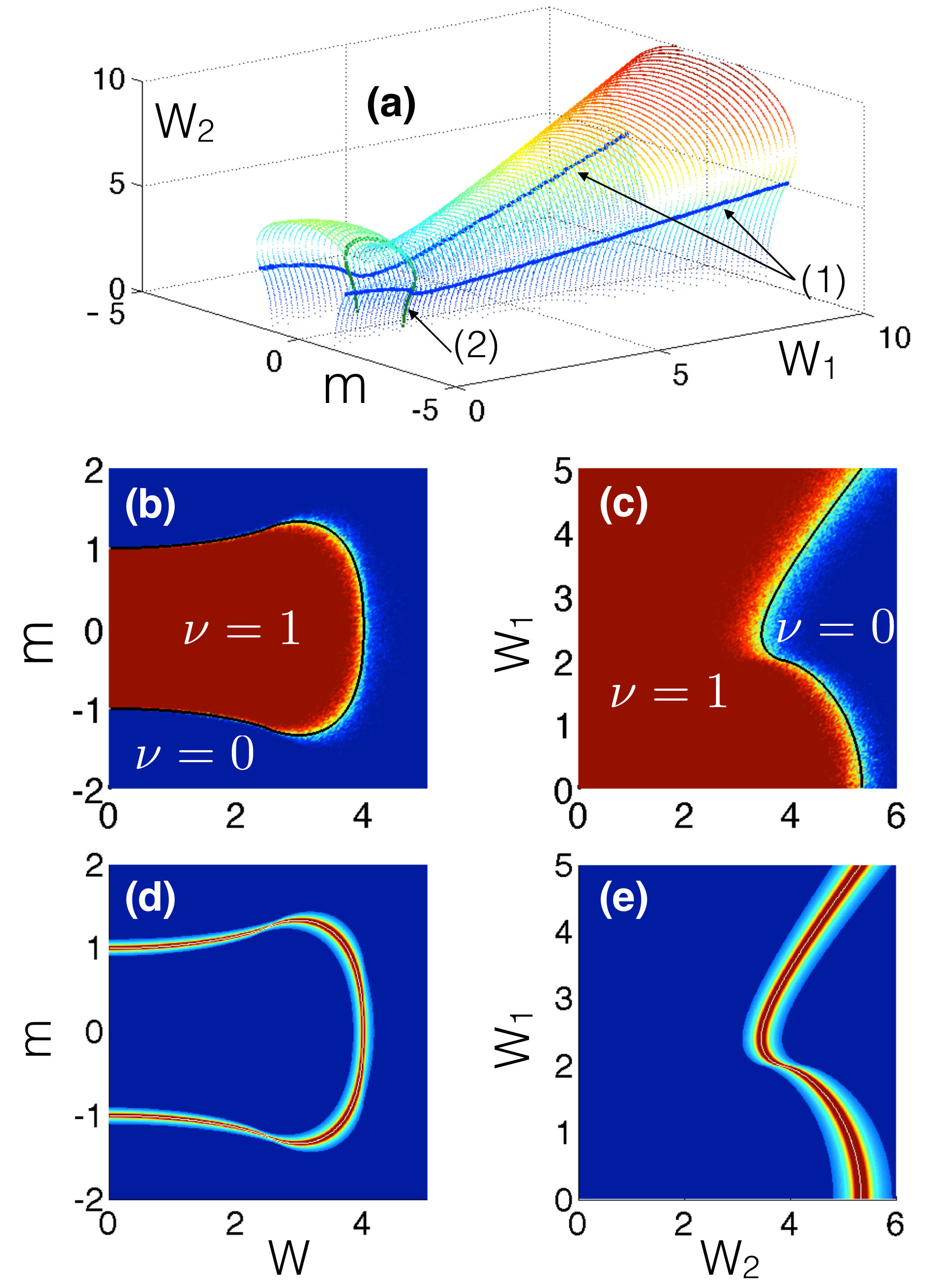}
\caption{ (a) The critical surface ($\Lambda \rightarrow \infty$) in the 3-dimensional phase space $(m,W_1,W_2)$. The lines (1) and (2) represent the singular points where the scaling is anomalous (see text). The next panels report maps of the winding number (b,c) and localization length (d,e) as computed with Eqs.~\ref{RWindingNr} and with the numerical transfer matrix method, respectively, for two sections of the phase space defined by the constraints $W_2=2W_1=W$ (b,d) and $m=0.5$ (c,e). The analytic critical curves are shown as black/white lines in panels (b,c)/(d,e), respectively. The computations of $\nu$ were done for $N=1000$  and averaged over 10 disorder configurations. The transfer matrix was iterated $10^8$ times.}\label{CriticalSurface}
\end{figure}

We begin by deriving a covariant, real-space representation of the winding number defined in Eq.~\ref{KWindingNr}. Such a real space representation is essential because it remains well-defined in the presence of disorder and can be evaluated with extreme precision using the methods elaborated in Refs.~\cite{Prodan2010ew,loring2010,ProdanJPhysA2011xk,hastings2011,ProdanAMRX2013bn}. For the derivation, it is more convenient to work with the homotopically equivalent flat band version of the Hamiltonian: $H \rightarrow Q \equiv P_+ - P_-$, where $P_\pm$ are the spectral projection operators onto the spectrum above/below $E=0.$ Since $S^\dagger = S$ and $S^2=1$, the eigenvalues of the chiral transformation are $\pm 1$, hence $S=S_+-S_-$, with $S_{\pm}$ being the projectors for the $\pm 1$ eigenvalues. Any chiral-symmetric operator, in particular Q, decomposes as: $Q=S_+ Q S_- + S_- Q S_+,$ and the following relations are always true:
$(S_\pm Q S_\mp)^\dagger = S_\mp Q S_\pm, \ \ (S_\pm Q S_\mp)^{-1}= S_\mp Q S_\pm.$
These provide the covariant, real-space form of the off-diagonal term (and its inverse) entering the winding number formula:
$Q_{+-}= S_+ Q S_-, \ (Q_{+-})^{-1}= S_- Q S_+=Q_{-+}.$
By recalling that $\int_0^{2\pi} \frac{dk}{2\pi}\mathrm{tr}\{A(k)\}$ for a generic translational invariant operator $A$ can be represented in real-space as a trace per volume (denoted by $\mathcal{T}$ here), and that $\partial_k$ is equal to the commutator $-i[X, \ ]$ in real-space ($X=$ the position operator), one can fully formulate the winding number in the real-space:
\begin{equation}\label{RWindingNr}
\nu =  -\ \mathcal{T}\big \{ Q_{-+} [X, Q_{+-}] \big \}.
\end{equation}
Following the same reasoning, one can derive a general covariant real-space representation of the invariant for the AIII-class in arbitrary $2n+1$-dimensions:
\begin{equation}\label{GRWindingNr}
\nu = \frac{-( \pi i)^n}{ (2n+1)!!} \sum_\rho (-1)^\rho \mathcal{T}\left \{ \prod_{i=1}^{2n+1}   Q_{-+} [X_{\rho_i}, Q_{+-}] \right \},
\end{equation}
where the summation is over all possible permutations $\rho$ of the indices. These real-space formulas can be evaluated in the presence of disorder, and is important to note that Eq.~\ref{GRWindingNr} is self-averaging, that is, the result of a computation is independent of the disorder configuration being used. For some disordered 1D topological phases one can formulate a topological invariant using transfer matrices as well\cite{Fulga2011}.

To be concrete, we fix $m=0.5$ in which case $\nu=1$ at $W_{1,2}=0$, and $\nu = 0$ in the limit $W_2 \rightarrow \infty$ because the onsite potential commutes with $X$. This signals a possible topological phase transition which we explored with Eq.~\ref{RWindingNr}. The behavior of $\nu$ with increasing disorder is reported in Fig.~\ref{InvariantVsW}. Here, one sees the winding number staying quantized and non-fluctuating even after the spectral gap closes from the strong disorder (no disorder averaging is necessary). Upon further increase of disorder, an abrupt switch occurs from the topological  $\nu=1$ to the trivial $\nu=0$ value, accompanied by strong fluctuations during the transition period. This behavior leaves little doubt that a topological critical point is lurking underneath.

 In order for $\nu$ to change values, there must be delocalized states appearing at the Fermi-level, which, for chiral symmetry, is at $E=0.$ Due to a simplification occurring precisely at $E=0$, the localization length of the disordered model and the critical exponents at the critical point can be computed exactly. Indeed, the Schrodinger equation $H\psi =0$ reads: $t_n\psi_{n-\alpha,\alpha}+i\alpha m_n \psi_{n,\alpha} =0$, where $\alpha = \pm1$ represents A and B site, respectively.
The solution is:
$$\psi_{n+\xi_\alpha,\alpha}=i^n \prod_{j=1}^{n} \left( \frac{t_j}{m_j}\right ) ^\alpha \psi_{\xi_\alpha,\alpha},$$
where $\xi_\alpha = 0,1$ for $\alpha=\pm 1$, respectively. The inverse of the localization length $\Lambda$ is given by:
$$
\begin{array}{l}
\Lambda^{-1} =\max_{\alpha=\pm 1}\big [- \lim\limits_{n\rightarrow \infty} \frac{1}{n}\log|\psi_{n+\xi_\alpha,\alpha}| \big ] \medskip \\
\indent =\big |\lim\limits_{n\rightarrow \infty} \frac{1}{n}\sum_{j=1}^{n}(\ln |t_j|-\ln|m_j|)\big |.
\end{array}
$$
According to Birkhoff's ergodic theorem, we can use the ensemble average to evaluate the last expression: 
$$
\begin{array}{l}
\Lambda^{-1} =\left | \ \int\limits_{-1/2}^{1/2}d\omega \int\limits_{-1/2}^{1/2}d\omega' \ (\ln|1+W_1 \omega| - \ln|m+W_2 \omega'|)\right |.
\end{array}
$$
The integrations can be performed explicitly, and in the regime of large $W$'s where the arguments of the logarithms can become negative, we obtain:
\begin{equation}\label{LocLength1}
\Lambda^{-1}=\left | \ln \left [\frac{|2+W_1|^{\frac{1}{W_1}+\frac{1}{2}}}{|2-W_1|^{\frac{1}{W_1}-\frac{1}{2}}}  \frac{| 2m-W_2|^{\frac{m}{W_2}-\frac{1}{2}}}{|2m+W_2|^{\frac{m}{W_2}+\frac{1}{2}}}  \right ] \right |.
\end{equation}
Using numerical transfer matrix and level-statistics analysis we combed the energy spectrum and found that, in every instance, all the states at $E \neq 0$ are localized. Hence, for the critical behavior, we can focus exclusively on the case $E=0.$ This enables us to use Eq. \ref{LocLength1} to draw the exact phase diagram in the 3-dimensional parameter space $(m, W_1, W_2)$ by tracing the critical surface $\mathcal{S}_c$ where $\Lambda \rightarrow \infty$. The result is shown in Fig.~\ref{CriticalSurface}(a), which reveals that we are indeed dealing with two phases that are completely disconnected from each other. We can show that the phase inside $\mathcal{S}_c$ is a topological phase with $\nu=1$ while outside $\mathcal{S}_c$, $\nu=0$.  As examples, in Figs.~\ref{CriticalSurface}(b,c) we show calculations of $\nu$ from Eq.~\ref{RWindingNr} for the sections defined by $W_2/W_1=2$ and $m=0.5$, respectively, which confirms that the topological phase $\nu=1$ extends all the way to the critical line, beyond which $\nu$ shifts abruptly to zero. By using the transfer matrix method \cite{MacKinnon1983}, the localization length was also obtained numerically in Figs. 2(d,e), where one can see a diverging critical line that matches perfectly the analytic critical line from Eq. (\ref{LocLength1}). Eq.~\ref{LocLength1} also enables us to determine the critical exponent for the transition. Let $(m_c,W_1^c,W_2^c)$ be a point on $\mathcal{S}_c$. We can cross $\mathcal{S}_c$ by varying any of the three parameters, so let us vary $m$ in a small interval $[m_c-\epsilon, m_c+\epsilon]$. As shown in the supplementary information $\Lambda^{-1}(m)=|m-m_c|[c_0+c_1(m-m_c)^2 \ldots]$, which gives a critical exponent 1, except along lines (1) and (2) shown in Fig. \ref{CriticalSurface}(a) where the scaling has a logarithmic correction $\Lambda^{-1}(m) \sim |m-m_c| \ln |m-m_c|.$

To find the physical origin of the topological phase transition, we map Eq.~\ref{Model} to a spin-$1/2$ Hamiltonian defined on a lattice of size $2N$ via the Jordan-Wigner transformation
\begin{eqnarray}
c_{n,A}=(-i)^{n+1} K(2n)S^{-}_{2n}, \quad c_{n,B}=(-i)^{n} K(2n-1) S^{-}_{2n-1}, \nonumber
\end{eqnarray}
where $S^a_{i}$ are spin-$1/2$ variables and $K(m)=\exp{\left(i\pi\sum_{j=1}^{m-1}S^{+}_j S^{-}_j\right)}$ is the kink operator. These transformations lead to a Hamiltonian
\begin{equation*}
H=\sum_{i }2t_i\left(\hat{S}^x_{2i} \hat{S}^x_{2i+1}+\hat{S}^y_{2i} \hat{S}^y_{2i+1}\right)+2m_{i}\left(\hat{S}^x_{2i} \hat{S}^x_{2i-1}+\hat{S}^y_{2i} \hat{S}^y_{2i-1}\right)
\end{equation*}
\noindent which is the spin-$1/2$ XX model with random exchange couplings $2t_i$ ($2m_{i+1}$) between the even (odd) bonds.
The simple form of the ground-state in the spin-representation can simplify the real-space description of the topological characterization. The ground state can be constructed by a real-space renormalization group (RG) procedure which is asymptotically exact\cite{Fisher1994,Dasgupta1980}. Each RG step consists of decimating the pair of spins that have the strongest exchange interaction by enforcing a spin-singlet state for that pair, and correspondingly generating a new and weaker bond between the neighboring spins. The final result is the ground state in which each spin forms a singlet state with another spin in the system.
For the generic case in which the distributions $t_i$ and $m_i$ are chosen to be different, the system is gapped and said to be dimerized\cite{Hyman1996}.  Roughly speaking, the topological and trivial phases correspond to dimerization patterns on either the odd or even bonds, and these patterns are preserved during each RG step.

 By the nature of the RG procedure, the singlets that are generated never cross each other, which implies that every singlet state in the ground state will be formed by one spin belonging to sublattice $A$ and another spin belonging to sublattice $B$ (a manifestation of the underlying chiral symmetry). Let us associate to the $i$-th singlet a pair of numbers $d_{i}=\{d_{i1},d_{i2}\}$ which are the lattice sites of the two spins in the singlet. The ground state is
\begin{eqnarray}
\ket{\Omega}&=&\prod_{i}\frac{1}{\sqrt{2}}\left[S^{\dagger}_{2d_{i1}}-S^{\dagger}_{2d_{i2}-1}\right]\ket{\downarrow \ldots \downarrow}.
\end{eqnarray}
If we map back to the fermion representation the ground state can be simplified to
$\ket{\Omega}=\prod_{i}\frac{1}{\sqrt{2}}\left[\alpha_i c^{\dagger}_{d_{i1},A}-\beta_ic^{\dagger}_{d_{i2},B}\right]\ket{0}$
where $ \alpha_{i}$ and $ \beta_{i}$ have unit modulus and depend on the configuration of the singlets (see supplementary information). This form of the ground state is remarkable because it is a product state constructed from the single-particle states $\ket{\Phi_i}=\frac{1}{\sqrt{2}}\left(\alpha_{i}c^{\dagger}_{d_{i1},A}-\beta_{i}c^{\dagger}_{d_{i2},B}\right)\ket{0}$
which, like the flat-band limit in the disorder-free system, only have weight on two sites (although now the sites can be far apart). In this basis, the real-space winding number formula drastically simplifies to
\begin{eqnarray}
\nu=\frac{1}{N}\sum_{i=1}^{N} (d_{i2}-d_{i1})\label{winding}
\end{eqnarray}
which is just the sum of vectors connecting the end-points of the singlets. In the clean topological phase $(d_{i2}-d_{i1})=1$ and $\nu=\frac{1}{N}\sum_{i}^N 1=1$, as expected. By contrast, in the trivial phase, singlets form onsite, which implies $\nu=\frac{1}{N}\sum_{i}^N 0=0$. This clearly illustrates that it is not a single de-localized state which carries the topological winding number, but instead, the entire set of occupied states. We could adiabatically deform the Hamiltonian, while preserving chiral symmetry, so that states of the form $\ket{\Phi_i}$ become the single-particle eigenstates and then \emph{each} state would carry a portion $\tfrac{1}{N}(d_{i2}-d_{i1})$ of $\nu.$


We can further exploit the mapping to the spin model to understand the nature of the topological phase transition. Consider disordering a state that is dimerized on the even bonds (i.e. the topological state). From the RG procedure, one can see that disorder will favour the formation of regions that dimerize on the odd bonds, so that trivial and topological regions coexist.  In the vicinity of the critical point, the low-energy interface states formed between these regions contribute to the spectral density inside the energy gap. This type of behavior corresponds to a Griffiths phase \cite{Motrunich2001,Hyman1996}, which is not critical and thus explains why the topological invariant does not change at the gap closing point. The system becomes critical when dimerization occurs equally on both the odd and even bonds, leading to a proliferation of zero-energy interface states. As a result, both the localization length as well as the density of states become divergent at zero energy [\onlinecite{Balents1997}]. Similar physics, albeit in a different context involving superconducting wires, was discussed in Refs. \onlinecite{Motrunich2001,Gruzberg2005}, both of which are important precursors for our work. The critical point realizes the random singlet (RS) phase in which singlets are formed on all length scales \cite{Fisher1995}. The divergent length scale of singlet formation would appear to destabilize the winding number form in Eq. \ref{winding} as expected at criticality. To further support the claim that the topological phase transition is in the same universality class as the RS phase we numerically confirmed (see supporting material) that the critical scaling of the entanglement entropy contains the $\log 2$ correction factor to the central charge as expected \cite{Refael2004}. Increasing the disorder beyond the critical point dimerizes the system on the odd bonds, which thus leads to the trivial state. 

In conclusion we have given a complete picture of the physics of the disordered AIII class in 1D. We have given a real-space formula for the AIII winding number in all odd dimensions, shown that topological invariants can be carried by localized states, and that, because of this, the levitation and annihilation topological phase transition is replaced by the random singlet transition for the AIII class in 1D. It will be exciting to see if these type of effects can be seen in higher dimensions.

\section*{ACKNOWLEDGMENTS}
IMS and TLH are supported by  ONR award N0014-12- 1-0935 and thank the UIUC ICMT for support. EP and JS acknowledge support by the U.S. NSF grants DMS-1066045, DMR-1056168. JS acknowledges additional support from NSFC under grants No. 11204065 and RFDPHE-China under Grant No. 20101303120005.
%

\clearpage

\onecolumngrid

\appendix

\section{Decomposition of Chiral-Symmetric operator}
Here we demonstrate that any operator $Q$ satisfying:
$$S Q S^{-1} = -Q, \ S^\dagger =S, \ S^2=1$$
can be written as:
$$Q=S_+ Q S_- + S_- Q S_+,$$
where $S=S_+ - S_-$ is the spectral decomposition of $S$. Noticing that:
$$Q = (S_+ + S_-) Q (S_+ + S_-),$$
we only need to show that $S_\pm Q S_\pm =0$. And indeed:
$$S_+ Q S_+ = -S_+ (SQS) S_+ = -(S_+S) Q (S S_+) = - S_+ Q S_+,$$
and similarly for $S_- Q S_-$. The affirmation then follows. Another immediate property is: 
$$ (S_\pm Q S_\mp) (S_\mp Q^{-1} S_\pm) = S_\pm,$$
which we also used in the text. The identity follows by observing that $ S_\pm Q S_\mp=S_\pm Q$ and that 
$S_\mp Q^{-1} S_\pm =  Q^{-1} S_\pm$, in which case:
$$ (S_\pm Q S_\mp) (S_\mp Q^{-1} S_\pm) = S_\pm Q Q^{-1} S_\pm= S_\pm.$$

\section{Proof of Polarization Relation}
Since ${\mathrm{Tr}}[A(k)]=i\sum_{\alpha\in occ.} \langle  u_\alpha(k)|\partial_k |u_\alpha(k)\rangle =i\sum_{\alpha\in occ.} \langle S u_\alpha(k)|\partial_k |S u_\alpha(k)\rangle$ the polarization of the occupied bands $P_o$ is equal to that of the unoccupied bands $P_u.$ Also, if we include all the bands then polarization is equal to an integer, and thus, we know that $P_o+P_u\in \mathbb{Z},$ and thus $2P_o\in \mathbb{Z}.$ This means that chiral symmetry quantizes the polarization in units of $1/2$, i.e., $P_o=\frac{n}{2}$ for $n\in \mathbb{Z}.$

Now let us consider a basis for a generic Bloch Hamiltonian $h(k)$ such that the chiral operator $S$ is diagonal. To be chiral symmetric, and gapped, $h(k)$ must have an even number of bands; half above zero energy and half below. Thus, when $S$ is diagonalized it generically takes the form $S=\tau^z\otimes \mathbb{I}_{N}$ where the total number of bands is $2N$, $\mathbb{I}_N$ is the $N\times N$ identity matrix, and $\tau^z$ is the diagonal Pauli matrix. With this choice of basis we can choose the Bloch functions of the occupied bands to be of the form
\begin{equation}
\vert u_{\alpha}(k)\rangle=\left(\begin{array}{cc} v_{1\alpha}(k)\\ v_{2\alpha}(k)\end{array}\right)
\end{equation}\noindent where $\alpha=1, 2, \ldots N$ and $v_{1\alpha}, v_{2\alpha}$ are $N$-component spinors. The unoccupied bands can then be written as
\begin{equation}
\vert S u_{\alpha}(k)\rangle=\left(\begin{array}{cc} v_{1\alpha}(k)\\ -v_{2\alpha}(k)\end{array}\right).
\end{equation}\noindent To satisfy normalization and the orthogonality of different bands we must have the constraints
\begin{eqnarray}\label{eq:polconstraint}
 v^{\dagger}_{1\alpha}(k)v_{1\beta}(k)+ v^{\dagger}_{2\alpha}(k)v_{2\beta}(k)&=&\delta_{\alpha\beta}\nonumber\\
v^{\dagger}_{1\alpha}(k)v_{1\beta}(k)- v^{\dagger}_{2\alpha}(k)v_{2\beta}(k)&=&0
\end{eqnarray}\noindent where the second constraint comes from $\langle Su_{\alpha}(k)\vert u_{\beta}(k)\rangle =0.$

Using this decomposition we can write twice the charge polarization of the occupied bands as
\begin{equation}
2P_o=\frac{2i}{2\pi}\int dk \sum_{\alpha\in occ.}\left[v^{\dagger}_{1\alpha}(k)\partial_k v_{1\alpha}(k)+ v^{\dagger}_{2\alpha}(k)\partial_k v_{2\alpha}(k)\right].
\end{equation} The winding number can be written
\begin{equation}
\nu=\frac{i}{\pi}\int dk \sum_{\alpha\in occ.}\left[v^{\dagger}_{1\alpha}(k)\partial_k v_{1\alpha}(k)- v^{\dagger}_{2\alpha}(k)\partial_k v_{2\alpha}(k)\right].
\end{equation}\noindent From the constraint that both $2P_o$ and $\nu$ are integers we know that $2P_o\pm \nu$ are also integers. This allows us to define
\begin{eqnarray}
2P_o+\nu\equiv c_{+}&=&\frac{4i}{2\pi}\int dk \sum_{\alpha\in occ.}\left[v^{\dagger}_{1\alpha}(k)\partial_k v_{1\alpha}(k)\right]\nonumber\\
2P_o-\nu\equiv c_{-}&=&\frac{4i}{2\pi}\int dk \sum_{\alpha\in occ.}\left[v^{\dagger}_{2\alpha}(k)\partial_k v_{2\alpha}(k)\right].
\end{eqnarray} From the definitions of $c_{\pm}$ we can see that
\begin{equation}
2P_o=\frac{c_{+}+c_{-}}{2},\;\;\;\; \nu= \frac{c_{+}-c_{-}}{2}.
\end{equation}\noindent For $2P_0, \nu$ to be integers $c_{\pm}$ must be both even or both odd. If they are both even then we have proven
\begin{equation}\label{eq:polfinal}
2P_o =\nu\mod 2
\end{equation}\noindent since in that case $2P_0=a+b$ and $\nu=a-b$ for integers $a,b$ and the sum and difference of two integers has the same parity. If both $c_{\pm}$ are odd then the result is
\begin{equation*}
2P_o=(\nu +1)\mod 2
\end{equation*}\noindent which means the polarization and winding have an opposite parity relationship. The final step is thus to prove that $c_{\pm}$ must always be even.

To show this we note that $\sqrt{2}v_{1\alpha}(k)$ and $\sqrt{2}v_{2\alpha}(k)$ are normalized to unity from the constraints in Eq. \ref{eq:polconstraint}. Thus, for each $\alpha$ we must have $\tfrac{i}{2\pi}\int dk\left[\sqrt{2}v^{\dagger}_{1\alpha}\partial_k \sqrt{2}v_{1\alpha}\right]$ and $\tfrac{i}{2\pi}\int dk\left[\sqrt{2}v^{\dagger}_{2\alpha}\partial_k \sqrt{2}v_{2\alpha}\right]$ both equal to integers. And finally, this shows that $c_{\pm}$ are even integers since they are sums over integers (one integer for each $\alpha$ that are subsequently multiplied by a global factor of two which makes the final result an even integer. This leaves us with Eq. \ref{eq:polfinal} as the correct relation between the polarization and winding number.

\section{Derivation of Critical Scaling}
Eq.~\ref{LocLength1} can also enable us to determine the critical exponent for the transition. Let $(m_c,W_1^c,W_2^c)$ be a point on $\mathcal{S}_c$. We can cross $\mathcal{S}_c$ by varying any of the three parameters, so let us vary $m$ in a small interval $[m_c-\epsilon, m_c+\epsilon].$
Note that Eq.~\ref{LocLength1} can rewritten as:
\begin{equation}\label{LocLength2}
\begin{array}{c}
\Lambda^{-1}(m)=\left | \ln \left [|2+W_1^c|^{\frac{1}{W_1^c}+\frac{1}{2}}\big /|2-W_1^c|^{\frac{1}{W_1^c}-\frac{1}{2}} \right ] \right .  \medskip \\
\left . + \left (\frac{m}{W_2^c}-\frac{1}{2}\right ) \ln | 2m-W_2^c| -\left (\frac{m}{W_2^c}+\frac{1}{2}\right ) \ln | 2m+W_2^c| \right |.
\end{array}
\end{equation}
One can see now explicitly that, except for the special cases when $m_c=\pm \nicefrac{1}{2}W_2^c$, the function inside the absolute value is analytic of m around $m_c$, and this function cancels when $m=m_c$. As such, $\Lambda^{-1}(m)=|m-m_c|[c_0+c_1(m-m_c)^2 \ldots]$, which gives a critical exponent 1. 

The special critical cases $m_c=\pm\nicefrac{1}{2}W_2^c$, marked with line (1) in Fig.~\ref{CriticalSurface}, occur when the amplitudes of the random and non-random components of the onsite potential are equal. If $m_c=\nicefrac{1}{2}W_2^c$:
\begin{equation}\label{LocLength2}
\begin{array}{c}
\Lambda^{-1}(m)=\left | \ln \left [|2+W_1^c|^{\frac{1}{W_1^c}+\frac{1}{2}}/|2-W_1^c|^{\frac{1}{W_1^c}-\frac{1}{2}} \right ] \right . \medskip \\
\left . + \frac{1}{W_2^c} (m-m_c ) \ln | 2(m-m_c)| -\frac{1}{W_2^c} (m+m_c) \ln | 2(m+m_c)| \right | .
\end{array}
\end{equation}
The function inside the absolute value sign cancels when $m=m_c$. Since the second term becomes null when $m=m_c$, the last term must cancel the first term when $m=m_c$, and this cancelation occurs in analytic fashion. As such, the behavior of $\Lambda^{-1}(m)$ around $m_c$ is 
$$\Lambda^{-1}(m)=\left | \frac{1}{W_2^c}|m-m_c| \ln |m-m_c| + c_1 |m-m_c| +c_2(m-m_c)^2 \ldots \right |,$$
hence the localization length diverges as:
$$ \Lambda(m) \sim \frac{W_2^c}{|m-m_c| \ln |m-m_c|}.$$
There is a second class of special critical points, given by $W_1^c=2$ and marked by line (2) in Fig.~\ref{CriticalSurface}, where a similar anomalous scaling occurs.

\section{Proof of the simplified form winding number from the spin representation}
As shown in the text, the ground state of the random spin-$1/2$ XX model is formed by singlets of varying lengths. Because of the no-crossing constraint on the singlets every singlet state in the ground state will be formed by one spin belonging to sublattice $A$ and another spin belonging to sublattice $B$. Let us associate to the $i$-th singlet the pair of numbers $d_{i}=\{d_{i1},d_{i2}\}$ which are the lattice sites of the two spins involved in forming the singlet. Using this notation, the ground state can be written as
\begin{eqnarray}
\ket{\Omega}&=&\prod_{i}\frac{1}{\sqrt{2}}\left[S^{\dagger}_{2d_{i1}}-S^{\dagger}_{2d_{i2}-1}\right]\ket{\downarrow \ldots \downarrow}\\
&=&\prod_{i}\frac{1}{\sqrt{2}}\left[(-i)^{d_{i1}+1}c^{\dagger}_{d_{i1},A} \tilde{K}(d_{i1},1)-(-i)^{d_{i2}}c^{\dagger}_{d_{i2},B}\tilde{K}(d_{i2},0)\right]\ket{0}\nonumber
\end{eqnarray}
where $\tilde{K}(m,\lambda)=e^{i\pi \lambda \hat{n}_{m,B}}\exp\left[i\pi\sum_{j=1}^{m-1} \left(\hat{n}_{jA}+\hat{n}_{jB}\right)\right]$. Due to the non-crossing nature of the singlets, the ground state can be simplified to
\begin{equation}
\ket{\Omega}=\prod_{i}\frac{1}{\sqrt{2}}\left[\alpha_i c^{\dagger}_{d_{i1},A}-\beta_ic^{\dagger}_{d_{i2},B}\right]\ket{0}
\end{equation}
where $ \alpha_{i}$ and $ \beta_{i}$ have unit modulus and their particular values depend on the configuration of the singlets.
Let us take this on faith for a moment, we will prove it below after we finish the proof of the simplified winding number formula in this representation of the ground state.

This form of the ground state is a Slater determinant constructed from the single-particle states
\begin{equation}
\ket{\psi_i}=\frac{1}{\sqrt{2}}\left(\alpha_{i}c^{\dagger}_{d_{i1},A}-\beta_{i}c^{\dagger}_{d_{i2},B}\right)\ket{0}
\end{equation}
which means that we can write explicitly $Q_{+-}=\sum_{i=1}^N\left( -\alpha_{i}\beta^*_i\right)\ket{d_{i1},A}\bra{d_{i2},B}$. The winding formula can thus be simplified in the following manner. First, we note that
\begin{eqnarray}
Q_{-+}\hat{X} Q_{+-}&=&\sum_{ij} \alpha^*_{i}\beta_i\alpha_{j}\beta^*_j\ket{d_{i2},B}\bra{d_{i1},A}\hat{X}\ket{d_{j1},A}\bra{d_{j2},B}\nonumber\\
 &=&\sum_{ij}\alpha^*_{i}\beta_i\alpha_{j}\beta^*_j\ket{d_{i2},B}d_{i1}\delta_{ij}\bra{d_{j2},B}\nonumber\\
&=&\sum_{i} d_{i1}\ket{d_{i2},B}\bra{d_{j2},B}.
\end{eqnarray}
Furthermore, we can write the position operator as $\hat{X}=\sum_{i} d_{i2}\ket{d_{i2},B}\bra{d_{j2},B}$, which means that we get
\begin{eqnarray}
-\left(Q_{-+}\hat{X} Q_{+-}-\hat{X}\right)=\sum_{i} (d_{i2}-d_{i1})\ket{d_{i2},B}\bra{d_{j2},B}.
\end{eqnarray}
Hence, the winding number can be explicitly written as
\begin{eqnarray}
\nu&=&-\mathcal{T}\left(Q_{-+}\left[\hat{X},Q_{+-}\right]\right)=-\mathcal{T}\left(Q_{-+}\hat{X} Q_{+-}-\hat{X}\right)\nonumber\\
&=&\frac{1}{N}\sum_{i=1}^{N} (d_{i2}-d_{i1})\label{windingApp}
\end{eqnarray}
This expression is just the sum of vectors connecting the end-points of the singlets. In particular, since the topological phase with no disorder corresponds to singlets forming between nearest neighbors, we have that $(d_{i2}-d_{i1})=1$. The invariant then yields $\nu=\frac{1}{N}\sum_{i}^N 1=1$, as expected. By contrast, in the trivial phase, singlets form between the $B$ and $A$ spins of the \textit{same} site, which implies $\nu=\frac{1}{N}\sum_{i}^N 0=0$.


To complete the proof we need to show that the singlet ground state can always be written in the form
\begin{equation}\label{simGSApp}
\ket{\Omega}=\prod_{i=1}^{N}\frac{1}{\sqrt{2}}\left[\alpha_i c^{\dagger}_{d_{i,1},A} -\beta_i c^{\dagger}_{d_{i,2},B}\right]\ket{0}
\end{equation}
where $\vert \alpha_i\vert=1$ and $ \vert \beta_i\vert=1$ for all $i$. We stress from the outset that the particular form of the $\alpha_i$ and $\beta_i$ is not important for the winding formula, since these factors drop out in the end. Thus, in the following derivation, we will be concerned exclusively with whether these factors have unit modulus.



We will prove Eq.~\ref{simGSApp} in an inductive manner. Let us define the following three operators which will help to simplify the notation:
\begin{eqnarray}
P(F,G)&=&\prod_{i=F}^{G}Z(i),\nonumber\\
Z(i)&=&\frac{1}{\sqrt{2}}\left[(-i)^{d_{i1}+1}c^{\dagger}_{d_{i1},A} \tilde{K}(d_{i1},1)-(-i)^{d_{i2}}c^{\dagger}_{d_{i2},B}\tilde{K}(d_{i2},0)\right],\nonumber\\
\tilde{Z}(i)&=&\frac{1}{\sqrt{2}}\left[\alpha_i c^{\dagger}_{d_{i1},A} -\beta_ic^{\dagger}_{d_{i2},B}\right]. \nonumber
\end{eqnarray}
The ground state, in terms of these operators, is given by $\ket{\Omega}=P(1,N)\ket{0}=\prod_{i=1}^{N}Z(i)\ket{0}$. The objective is then to show that $\ket{\Omega}=\prod_{i}^{N}\tilde{Z}(i)\ket{0}$. To simplify the derivation, we will assume that the labeling of the singlets is such that they are organized according to their length, so that the smallest singlet acts on the vacuum first. This can always be done because the factors in $\ket{\Omega}$ (previous to any simplification) all commute. We start by rewriting the ground state as
\begin{eqnarray}
\ket{\Omega}&=&
P(1,N-1)Z(N)\ket{0} \nonumber\\
&=&
P(1,N-1)\frac{1}{\sqrt{2}}\left[(-i)^{d_{N,1}+1}c^{\dagger}_{d_{N,1},A} -(-i)^{d_{N,2}}c^{\dagger}_{d_{N,2},B}\right]\ket{0} \nonumber\\
&=&P(1,N-1)\tilde{Z}(N)\ket{0} \nonumber
\end{eqnarray}
where we have used that $\tilde{K}(m,\lambda)\ket{0}=\ket{0}$. We now simplify the next factor:
\begin{eqnarray}
\ket{\Omega}&=&P(1,N-1)\tilde{Z}(N)\ket{0}\\
&=&P(1,N-2)\frac{1}{\sqrt{2}}\left[(-i)^{d_{N-1,1}+1}c^{\dagger}_{d_{N-1,1},A}\tilde{K}(d_{N-1,1},1) -(-i)^{d_{N-1,2}}c^{\dagger}_{d_{N-1,2},B}\tilde{K}(d_{N-1,2},0)\right]\tilde{Z}(N)\ket{0} \\
&=&P(1,N-2)\frac{1}{\sqrt{2}}\left[(-i)^{d_{N-1,1}+1}c^{\dagger}_{d_{N-1,1},A}\tilde{K}(d_{N-1,1},1) \tilde{Z}(N)\ket{0} -(-i)^{d_{N-1,2}}c^{\dagger}_{d_{N-1,2},B}\tilde{K}(d_{N-1,2},0)\tilde{Z}(N)\ket{0} \right].
\end{eqnarray}
Because of the non-crossing rule and the way we have organized the singlets according to their length, there are three possible situations:
\begin{enumerate}
\item \textit{The interval $d_{N}$ lies to the right of the interval $d_{N-1}$:} In this case, neither $\tilde{K}(d_{N-1,2},0)$ nor $\tilde{K}(d_{N-1,1},1)$ will have number operators in their exponentials with corresponding creation operators in $Z(N)$, which means that the $K$ operators commute with $Z(N)$. Hence, we can write $\ket{\Omega}=P(1,N-2)\tilde{Z}(N-1)\tilde{Z}(N)\ket{0}$.
\item \textit{The interval $d_{N}$ lies inside of the interval $d_{N-1}$:} Then one, and only one, of $\tilde{K}(d_{N-1,2},0)$ and $\tilde{K}(d_{N-1,1},1)$ will have the number operators associated with \textit{both} of the creation operators in $Z(N)$. The state $Z(N)\ket{0}$ is thus an eigenstate of both $K$ operators, with eigenvalues $\pm1$, depending on which $K$ has operators shared with $Z(N)$. Hence, $\ket{\Omega}=P(1,N-2)\tilde{Z}(N-1)\tilde{Z}(N)\ket{0}$.
\item \textit{The interval $d_{N}$ lies to the left of $d_{N-1}$:} Then both $\tilde{K}(d_{N-1,2},0)$ and $\tilde{K}(d_{N-1,1},1)$ will have the number operators associated with \textit{both} of the creation operators in $Z(N)$. The state $Z(N)\ket{0}$ is thus an eigenstate of both $K$ operators, with eigenvalues $-1$. Hence, $\ket{\Omega}=P(1,N-2)\tilde{Z}(N-1)\tilde{Z}(N)\ket{0}$.
\end{enumerate}
In all three cases, we get the desired simplification of the ground state. This is the first inductive step. For the $m$-th step of the inductive argument, we now assume that the following expression holds
\begin{equation}
\ket{\Omega}=P(1,N-m)\prod_{i=N}^{N-m+1}\tilde{Z}(i)\ket{0},
\end{equation}
and we must now show that this implies the $m+1$ case, namely that
\begin{equation}
\ket{\Omega}=P(1,N-m-1)\prod_{i=N}^{N-m}\tilde{Z}(i)\ket{0}.
\end{equation}
To show this, we again write
\begin{eqnarray}
\ket{\Omega}&=&P(1,N-m)\prod_{i=N}^{N-m+1}\tilde{Z}(i)\ket{0}\\
&=&P(1,N-m-1)\frac{1}{\sqrt{2}}\left[(-i)^{d_{N-m,1}+1}c^{\dagger}_{d_{N-m,1},A}\tilde{K}(d_{N-m,1},1) -(-i)^{d_{N-m,2}}c^{\dagger}_{d_{N-m,2},B}\tilde{K}(d_{N-m,2},0)\right]\prod_{i=N}^{N-m+1}\tilde{Z}(i)\ket{0} 
\end{eqnarray}
We can group the factors $\prod_{i=N}^{N-m+1}\tilde{Z}(i)=B_1 B_2 B_3$ in the following way:
\begin{itemize}
\item \textit{$B_1$ is made out of singlets with intervals that lie to the right of the interval $d_{N-m}$:} In this case, neither $\tilde{K}(d_{N-m,2},0)$ nor $\tilde{K}(d_{N-m,2},1)$ will have number operators in their exponentials with corresponding creation operators in $B_1$, which means that the $\tilde{K}$ operators commute with $B_1$. 
\item \textit{$B_2$ is made out of singlets with intervals that lie inside of the interval $d_{N-m}$:} Then one, and only one, of $\tilde{K}(d_{N-m,2},0)$ and $\tilde{K}(d_{N-m,2},1)$ will have \textit{all} of the number operators associated with the creation operators in $B_2$. Let us denote the number of factors in $B_2$ by $\rho_{B_2}$.
\item \textit{$B_3$ is made out of singlets with intervals such that $d_{N-m}$ lies to the left of them:} Then both $\tilde{K}(d_{N-m,2},0)$ and $\tilde{K}(d_{N-m1,2},1)$ will have \textit{all} of the number operators associated with the creation operators in $B_3$. Let us denote the number of factors in $B_2$ by $\rho_{B_3}$.
\end{itemize}
Hence, depending on which case occurs with the $B_2$ singlets, we will either get
\begin{eqnarray}
\ket{\Omega}&=&P(1,N-m-1)\frac{(-1)^{n_{B_3}}}{\sqrt{2}}\left[(-i)^{d_{N-m,1}+1}c^{\dagger}_{d_{N-m,1},A}  (-1)^{n_{B_2}}-(-i)^{d_{N-m,2}}c^{\dagger}_{d_{N-m,2},B}\right]B_1 B_2 B_3\ket{0}
\end{eqnarray}
or
\begin{eqnarray}
\ket{\Omega}&=&P(1,N-m-1)\frac{(-1)^{n_{B_3}}}{\sqrt{2}}\left[(-i)^{d_{N-m,1}+1}c^{\dagger}_{d_{N-m,1},A} -(-i)^{d_{N-m,2}}c^{\dagger}_{d_{N-m,2},B} (-1)^{n_{B_2}}\right]B_1 B_2 B_3\ket{0}
\end{eqnarray}
In both cases, we conclude that the singlet ground state can be written as
\begin{equation}
\ket{\Omega}=P(1,N-m-1)\prod_{i=N}^{N-m}\tilde{Z}(i)\ket{0}.
\end{equation}
As a consequence, by successive application of this procedure, we can finally arrive at the desired result
\begin{equation}
\ket{\Omega}=\prod_{i=1}^{N}\tilde{Z}(i)\ket{0}.
\end{equation}

\begin{figure}
\includegraphics[scale=0.4]{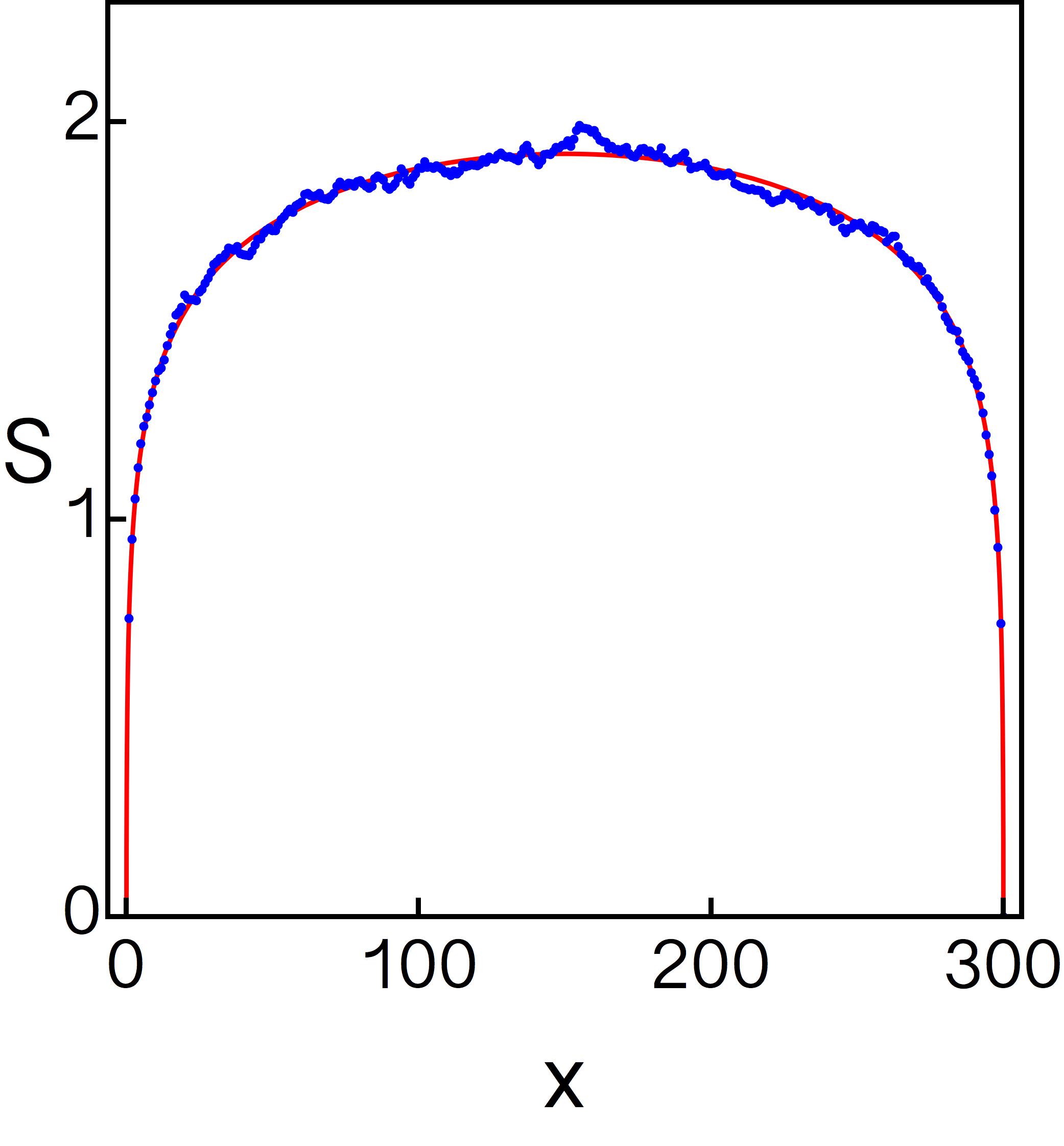}
\caption{ Scaling of the entanglement entropy of a section of size $x$ when the parameters are tuned to criticality. The blue dots correspond to the numerical calculation, and averaging was done over $200$ disorder realizations for a lattice of size $N=300$. The red solid line corresponds to Eq.~\ref{scal}.}\label{fig:entanglement}
\end{figure}

\section{Scaling of Entanglement Entropy}
Considering the scaling of the entanglement entropy as a function of the partition size has become a common technique for the study of 1d critical points\cite{calabrese2004,refael2009}. When tuned to the critical point one expects that the entanglement entropy will scale proportional to the $\log$ of the partition size with a proportionality constant that includes the central charge of the 1d CFT critical point. For the random-singlet critical point the scaling should be
\begin{equation}
S=\frac{c \log(2)}{3} \log\left[\frac{N}{\pi} \sin\left(\frac{\pi x}{N}\right)\right]+s_0 \label{scal}
\end{equation}
where $s_0$ is a non-universal constant and $c$ is the central charge of the CFT associated with the system at criticality in the clean limit, which is $c=1$ in our case, $N$ is the total length of the system and $x$ is the length of the partition size. The $\log(2)$ factor is a disorder-induced ``renormalization''  of the central charge that is characteristic for certain disordered spin models \cite{Refael2004}. We explored the scaling of the entanglement using the fermionic model at the critical point, and found that it matches Eq.~\ref{scal}. This further supports the notion that the critical point of the chiral symmetric model is  that of the random singlet phase. The numerical fit can be found in Fig. \ref{fig:entanglement}.


\end{document}